\documentclass[sigconf]{acmart}

\usepackage{wrapfig}
\usepackage{graphicx}
\usepackage{url}
\usepackage{multicol} 
\usepackage{flushend}

\copyrightyear{2019}
\acmYear{2019}
\setcopyright{acmcopyright}
\acmConference[HoTSoS '19]{Hot Topics in the Science of Security:
Symposium and Bootcamp}{April 2--3, 2019}{Nashville, TN, USA}
\acmBooktitle{HoTSoS '19: Hot Topics in the Science of Security: Symposium
and Bootcamp, April 2--3, 2019, Nashville, TN, USA}

\begin{document}
\title{Game Theory for Cyber Deception}
\subtitle{A Tutorial}

\author{Quanyan Zhu}
\affiliation{%
  \institution{New York University}
  \streetaddress{5 MetroTech Center}
  \city{Brooklyn}
  \state{New York}
  \postcode{11201}
}
\email{qz494@nyu.edu}


\renewcommand{\shortauthors}{Q. Zhu}

\begin{abstract}

Deceptive and anti-deceptive technologies have been developed for various specific applications. But there is a significant need for a general, holistic, and quantitative framework of deception. Game theory provides an ideal set of tools to develop such a framework of deception. In particular, game theory captures the strategic and self-interested nature of attackers and defenders in cybersecurity. Additionally, control theory can be used to quantify the physical impact of attack and defense strategies. In this tutorial, we present an overview of game-theoretic models and design mechanisms for deception and counter-deception. The tutorial aims to provide a taxonomy of deception and counter-deception and understand how they can be conceptualized, quantified, and designed or mitigated. This tutorial gives an overview of diverse methodologies from game theory that includes games of incomplete information, dynamic games, mechanism design theory to offer a modern theoretic underpinning of cyberdeception.  The tutorial will also discuss open problems and research challenges that the HoTSoS community can address and contribute with an objective to build a multidisciplinary bridge between cybersecurity, economics, game and decision theory.


\end{abstract}

%
%



\maketitle


\section*{Tutorial Description}

Cyber deception is a technique used to cause human \cite{vrij2008increasing, gneezy2005deception} or computer systems \cite{bodmer2012reverse,horak2017manipulating,Huang2019} to have false beliefs and behave against their interests. %
Cyberspace creates unique opportunities for deception. 
Attackers can leverage information asymmetry to misinform and mislead the users. For example, phishing is a typical deception-based attack that is one of the top threat vectors for cyberattacks. 
Defenders can also use deception as a way to thwart or deter attacks. For example, techniques such as honeynets \cite{carroll2011game,zhu2013deployment}, moving target
defense \cite{zhu2013game,jajodia2011moving,huang2019adaptive}, obfuscation \cite{pawlick2016stackelberg,zhang_dynamic_2017,farhang2015phy,zhang2018distributed},
and mix networks \cite{zhang2010gpath} have been introduced to create difficulties for attackers to map out the system information.

Successful deception fundamentally depends on the information asymmetry between the deceiver and the deceivee \cite{zhang2019game,zhu2018gameTutorial,Pawlick2018Dissertation,pawlick2017game}.  Deceivers can strategically manipulate the private information to suit their own self-interests. The manipulated information is then revealed to deceivees, who, on the other hand, make decisions based on the information received. 
It is important for the deceivee to form correct beliefs based on past observations, take into account the potential damage caused by deception, and strategically use the observed information for decision-making. 
%

Modeling deception would formally capture the  interactions between an attacker and a defender, provide a quantitative and systematic understanding of deceptions, and develop new technologies and incentive mechanisms to mitigate cyber risks and safeguard the cyberspace. Game-theoretic models are natural frameworks to capture the adversarial and defensive interactions between players \cite{manshaei2013game}. It provides a quantitative measure of the quality of protection with the concept of Nash equilibrium where both defender and an attacker seek optimal strategies, and no one has an incentive to deviate unilaterally from their equilibrium strategies despite their conflict for security objectives. The equilibrium concept also provides a quantitative prediction of the security outcomes of the scenario the game model captures. Recently, game theory has been applied to  different sets of
security problems, e.g., Stackelberg and signaling games for
deception and proactive defenses
\cite{pawlick_stackelberg_2016,zhu2013game,zhu2013deployment,zhu2013hybrid,zhu2012interference,clark2012deceptive,zhu2012game,zhu2012deceptive,zhu2010stochastic}, (2) network games for cyber-physical security that deals with critical infrastructure protection and information assurance
\cite{xu2017secure,xu_game-theoretic_2017,xu_cross-layer_2016,farooq2019modeling,xu2015cyber,huang2017large,chen2017dynamic,miao2018hybrid,yuan2013resilient,Rass&Zhu2016,Rass.2017b},
 dynamic games for adaptive defense
\cite{zhu2010dynamic,zhang2017strategic,huang2018gamesec,huang2018PER,huang2019adaptive,pawlick2015flip,farhang2014dynamic,zhu2009dynamic,zhu2010network,zhu2010heterogeneous}, and mechanism design theory for security
\cite{chen_security_2017,zhang_bi-level_2017,zhang_attack-aware_2016,casey2015compliance,hayel2015attack,hayel2017epidemic,zhu2012guidex,zhu2012tragedy,zhu2009game}.
%

This tutorial aims to present an overview of game-theoretic methods and show their applications in different scenarios. More specifically, we start with a baseline a multi-stage Bayesian game model with two-sided incomplete information and introduce the analysis of different variants of the baseline that can be used to capture different features of cyber deception. The objective of this tutorial is to introduce diverse methodologies from
game theory that include mechanism design, incentive analysis,
decision-making under incomplete information, and dynamic games to provide
solid underpinnings of cyber deception. %

This tutorial will be structured in the following way. The first part of the tutorial introduces a taxonomy of deception from a game-theoretic perspective \cite{pawlick2017game,Pawlick2018Dissertation}. Then, the tutorial introduces the baseline signal games and the analytical methods for two-person deception games \cite{pawlick2015deception,casey_compliance_2016,pawlick2018modeling,zhang2017strategic}. We will present the concept of strategic trust as a defense mechanism that can be designed using the signaling games \cite{pawlick2017strategic,pawlick2018istrict,horak2017manipulating,xu2015cyber,moghaddam2015trust,fung2016facid}.  The baseline games can be further extended to include side-channel information, system dynamics, and the influence of the third party.  The tutorial will also discuss open problems and research challenges that the HoTSoS community can address and contribute. With the objective to build a multidisciplinary bridge between cybersecurity, economics, game and decision theory, this tutorial will review basic concepts and provide an overview of recent advances in the field to HoTSoS community with the hope to establish a community interest in the science of security and cross-disciplinary researches.

The potential audience includes researchers from academia and industry, including PhD and graduate students. Some background in network security and knowledge of basic optimization and data science is helpful but not necessary. 

\section{Author Biography}

\begin{wrapfigure}{r}{0.14\textwidth}
  \begin{center}
  \vspace{-4mm}
   \includegraphics[width=0.13\textwidth]{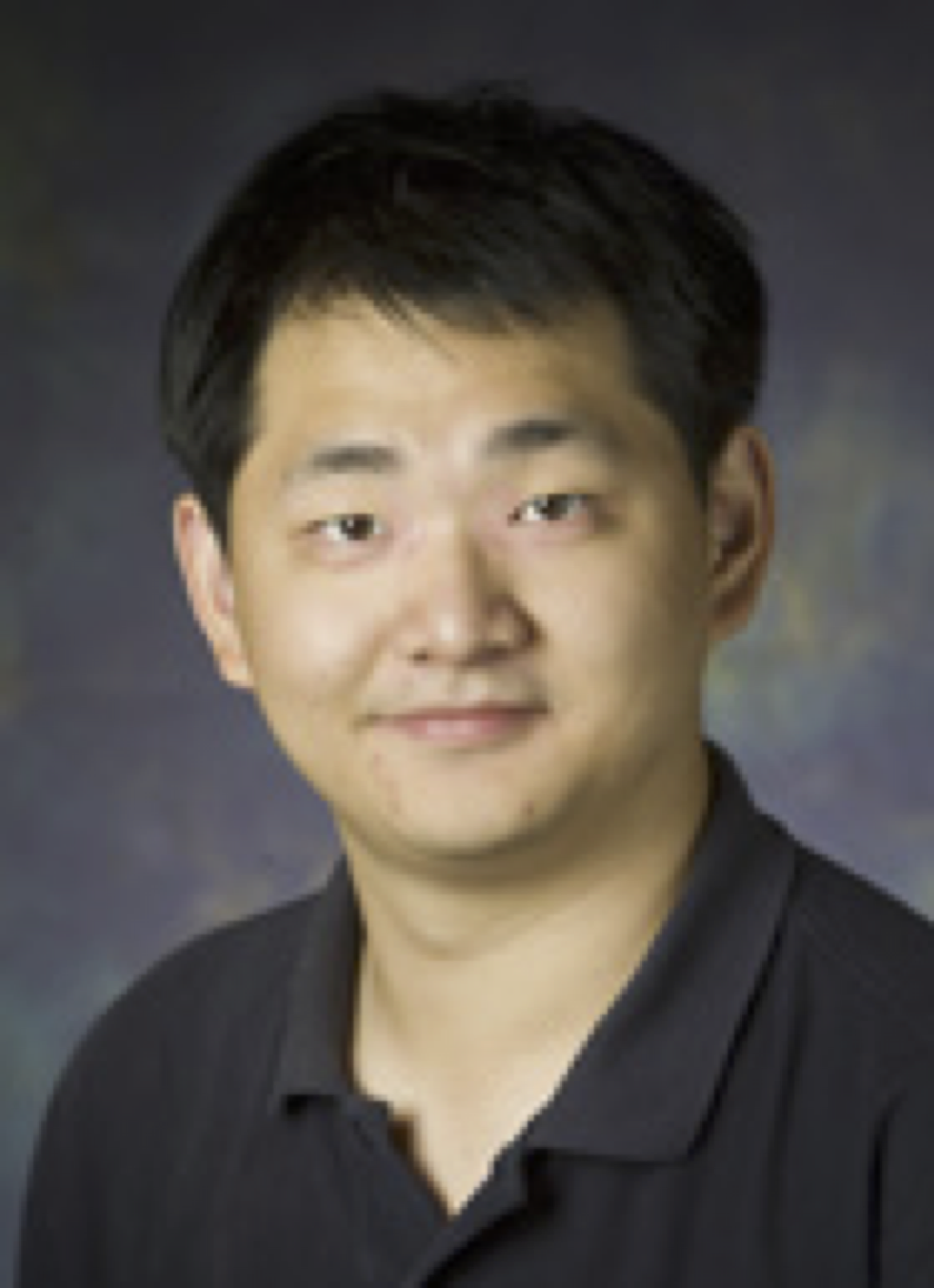}
  \end{center}  
\end{wrapfigure}


Quanyan Zhu received B. Eng. in Honors Electrical Engineering with distinction from McGill University in 2006, M.A.Sc. from University of Toronto in 2008, and Ph.D. from the University of Illinois at Urbana-Champaign (UIUC) in 2013. After stints at Princeton University, he is currently an assistant professor at the Department of Electrical and Computer Engineering, New York University. He is a recipient of many awards including NSF CAREER Award, NYU Goddard Junior Faculty Fellowship, NSERC Postdoctoral Fellowship (PDF), NSERC Canada Graduate Scholarship (CGS), and Mavis Future Faculty Fellowships. He spearheaded and chaired INFOCOM Workshop on Communications and Control on Smart Energy Systems (CCSES), and Midwest Workshop on Control and Game Theory (WCGT). His current research interests include resilient and secure interdependent critical infrastructures, Internet of Things, cyber-physical systems, game theory, machine learning, network optimization and control.  He is a recipient of best paper awards at 5th International Conference on Resilient Control Systems and 18th International Conference on Information Fusion. He has served as the general chair of the 7th Conference on Decision and Game Theory for Security (GameSec) in 2016, the 9th International Conference on NETwork Games, COntrol and OPtimisation (NETGCOOP) in 2018, and the 5th International Conference on Artificial Intelligence and Security (ICAIS 2019) in 2019. Website: \url{http://wp.nyu.edu/quanyan}

%
%
%

\bibliographystyle{acm}

\bibliography{MyPapers-BIBTEX-2018,additional-refs,reference}

\end{document}